\documentclass[12pt]{article}
\begin{document}

\centerline{\bf Causality in the brane world}

\bigskip

\centerline{Philip D. Mannheim}
\centerline{Department of Physics,
University of Connecticut, Storrs, CT 06269, USA}
\centerline{philip.mannheim@uconn.edu}
\medskip
\centerline{July 6, 2006}

\begin{abstract}

For the Randall-Sundrum brane world where a positive tension Minkowski
brane is embedded in $AdS_5$, two candidate propagators have been
suggested in the literature, one being based on the normalized mode
solutions to the source-free volcano potential fluctuation equation, and
the other being the Giddings, Katz and Randall outgoing Hankel function
based one. We show that while both of these two propagators have the same
pole plus cut singularity structure in the complex energy plane, they
behave differently on their respective circles at infinity, as a
consequence of which only the Hankel function based propagator proves to
be causal, with the normalized mode based one being found to take support
outside the $AdS_5$ lightcone. In addition we show that unlike the Hankel
function based propagator, the normalized mode based propagator does not
correctly implement the junction conditions which hold in the presence of
a perturbative source on the brane.

\end{abstract}

\section{The normalized mode propagator}

The Randall-Sundrum brane world associated with the embedding of a
positive tension $M_4$ Minkowski brane in an $AdS_5$ space with constant
negative curvature $-b^2$ can be characterized by a background metric
given as
\cite{Randall1999}
\begin{equation}
ds^2=dw^2+e^{2A(|w|)}(dx^2+dy^2+dz^2-dt^2)
\label{1.1}
\end{equation}
where $A=-b|w|$. (Here $w$ denotes the fifth coordinate which is to
accompany the familiar $x,y,z,t$, with the brane being at $w=0$.) In the
brane world one is interested in gravitational fluctuations around this
background, with the axial gauge, transverse-traceless ($TT$) ones being
found (see e.g. \cite{Mannheim2005} where full bibliographical citations
as well as details of the present work are given) to obey the wave
equation
\begin{equation}
{1 \over 2}\left[{\partial^2 \over \partial w^2}
-4\left({d A \over d|w|}\right)^2
-4{d A\over d|w|}\delta(w)
+e^{-2A}\eta^{\alpha\beta}\partial_{\alpha}\partial_{\beta}
\right]h^{^{TT}}_{\mu\nu}(x,|w|)
=-\kappa_5^2\delta(w)S^{^{TT}}_{\mu\nu}(x)
\label{1.2}
\end{equation}
when a perturbative source $S_{\mu\nu}(x)$ is placed on the
brane \cite{footnote0}.  To integrate a wave
equation one ordinarily constructs a propagator using a set of basis
modes which obey the source-free variant of the wave equation, and on
separating Eq. (\ref{1.2}) via
$\eta^{\alpha\beta}\partial_{\alpha}\partial_{\beta}h^{^{TT}}_{\mu\nu}=
m^2h^{^{TT}}_{\mu\nu}$, one thus sets
$h_{\mu\nu}^{^{TT}}=f_m(|w|)e^{^{TT}}_{\mu\nu}(x^{\lambda},m)$ where the 
$f_m(|w|)$ modes obey
\begin{equation}
\left[{d^2 \over d |w|^2}
-4\left({d A\over
d|w|}\right)^2
+e^{-2A}m^2
\right]f_m(|w|) =0
\label{1.3}
\end{equation}
\begin{equation}
\delta(w)\left[{d \over
d|w|}-2{d A\over
d|w|}\right]f_m(|w|)=0~~.
\label{1.4}
\end{equation}
These $f_m(|w|)$ modes are commonly called volcano potential modes
(because of the shape of the potential in Eq. (\ref{1.3})), with
manipulation of Eqs. (\ref{1.3}) and Eq. (\ref{1.4}) showing that every
pair of such modes obeys  
\begin{equation}
(m_1^2-m_2^2)\int_0^{\infty}d|w|e^{-2A}f_{m_1}f_{m_2}=
\lim_{|w| \rightarrow \infty}
\left[f_{m_1}\left({d \over d|w|}-2{dA \over d|w|}\right)f_{m_2}
-f_{m_2}\left({d \over d|w|}-2{dA \over d|w|}\right)f_{m_1}\right]~~.
\label{1.5}
\end{equation}
The requirement of asymptotic vanishing of the modes as $|w| \rightarrow 
\infty$ restricts the modes to ones which then obey the
orthonormality and closure relations
\begin{equation}
\int^{\infty}_{-\infty}dw e^{-2A}f_{m}(|w|)f_{m^{\prime}}(|w|)
=\delta_{m,m^{\prime}}~~,~~
\sum_{m} f_m(|w|)f_{m}(|w^{\prime}|)
=e^{2A}\delta(w-w^{\prime})~~, 
\label{1.6}
\end{equation}
with the normalized mode propagator \cite{Garriga2000} 
\begin{eqnarray}
G^{{\rm NM}}(x,x^{\prime},w,w^{\prime})&=&
\sum_{m}f_m(|w|)
f_m(|w^{\prime}|)
D(x-x^{\prime},m)
\nonumber \\
&=&\sum_{m}f_m(|w|)
f_m(|w^{\prime}|)
\int {d^4p \over (2\pi)^4}{e^{ip\cdot (x-x^{\prime})} \over
[(p^0)^2-\bar{p}^2-m^2+i\epsilon \epsilon(p^0)]}
\label{1.7}
\end{eqnarray}
[$D(x-x^{\prime},m)$ being the standard flat $M_4$
space retarded propagator which obeys
$[\eta^{\alpha\beta}\partial_{\alpha}\partial_{\beta}-m^2]
D(x-x^{\prime},m)=\delta^4(x-x^{\prime})$]
then serving as a propagator with which to integrate Eq. (\ref{1.2})
according to 
\begin{equation}
h^{^{TT}}_{\mu\nu}(x,|w|;{\rm NM})=-2\kappa_5^2\int d^4x^{\prime}G^{{\rm
NM}}(x, x^{\prime},w,0) S^{^{TT}}_{\mu\nu}(x^{\prime})~~.
\label{1.8}
\end{equation}

For $A=-b|w|$ the explicit basis modes associated with Eqs.
(\ref{1.3}) and Eq. (\ref{1.4}) are given as an $m^2=0$ massless graviton
with wave function $f_0(y)=\alpha_0 e^{-2b|w|}$, and an $m^2>0$ KK
continuum with $f_m(y)=\alpha_m J_{2}(y)+\beta_m Y_{2}(y)$ (here
$y=me^{b|w|}/b$) as constrained according to $\alpha_m J_1(m/b)+\beta_m
Y_1(m/b)=0$ \cite{footnote1}. After appropriately normalizing the
$f_m(|w|)$ modes \cite{footnote2}, the $p^0$ contour integration can be
performed in Eq. (\ref{1.7}), and with the retarded propagator contour
putting all singularities below the real $p^0$ axis, closing the contour
below the real axis yields a singular contribution to
$G^{{\rm NM}}(x,0,w,0)$ of the form 
\begin{eqnarray}
&&G^{{\rm NM}}(x,0,w,0; {\rm SING})=
-ibe^{-2b|w|}\int {d^3p\over (2\pi)^3}
 {e^{i\bar{p}\cdot \bar{x}}
\over 2|p|}\left[e^{-i|p|t}-e^{i|p|t}\right]
\nonumber \\
&&~~~~~-i\sum_m {b[Y_1(m/b)J_2(me^{b|w|}/b)-J_1(m/b)Y_2(me^{b|w|}/b)]
\over
\pi[J_1^2(m/b)+Y_1^2(m/b)]}
\int{d^3p  \over (2\pi)^3}
{e^{i\bar{p}\cdot \bar{x}}
\over
2E_p}\left[e^{-iE_pt}
-e^{iE_pt}\right]~~,
\label{1.9}
\end{eqnarray}
together with a contribution due to the complex $p^0$ lower half 
plane circle at infinity which is of the form of (the negative of)
\begin{eqnarray}
G^{{\rm NM}}(x,0,w,0;{\rm LHPC})&=&
\sum_{m}f_m(|w|)
f_m(0)
\nonumber \\
&&\times\left[-{1\over 4\pi
r}\delta(t+r)+\frac{m}{4\pi(t^2-r^2)^{1/2}}\theta(-t-r)
J_1\left(m(t^2-r^2)^{1/2}\right)\right]~~.
\label{1.10}
\end{eqnarray}
With $G^{{\rm NM}}(x,0,w,0;{\rm LHPC})$ vanishing when $t$ is positive,
the complete $t>0$ normalized mode propagator $G^{{\rm NM}}(t>0,x=0, y=0,
z=0,0,w,0)$ is thus given entirely via its singular part. And thus to
check for causality we need to determine whether or not $G^{{\rm
NM}}(t>0,x=0, y=0, z=0,0,w,0;{\rm SING})$ takes support outside  the
$AdS_5$ lightcone. Thus with Eq. (\ref{1.1}) entailing that the $AdS_5$
lightcone and its interior are given by 
\begin{equation}
\alpha={1 \over b}(bt-e^{b|w|}+1) \geq 0
\label{1.11}
\end{equation}
for the relevant points of interest, we thus need to determine whether or
not $G^{{\rm NM}}(t>0,x=y=z=0,x^{\prime}=0,w,w^{\prime}=0;{\rm SING})$
vanishes when
$\alpha$ is negative. And since the expression for $G^{{\rm NM}}(x,
x^{\prime},w,w^{\prime})$ given in Eq. (\ref{1.7}) involves a
direct product of functions of $|w|$ with the propagator
$D(x-x^{\prime},m)$ (a propagator which knows only about causality in
$M_4$) while not containing any immediately apparent $\theta(\alpha)$ type
dependence, we
shall anticipate, and in shall in fact shortly show, that the $G^{{\rm
NM}}(x,0,w,0)$ propagator is not actually causal in the full $AdS_5$
space. 

\section{The Giddings, Katz and Randall propagator}

In order to address the causality issue, it is convenient to introduce
an alternate brane-world propagator, the Giddings, Katz and
Randall one \cite{Giddings2000}, which enables us to integrate Eq.
(\ref{1.2}) according to 
\begin{eqnarray}
h^{^{TT}}_{\mu\nu;}(x,|w|;{\rm GKR})&&=-{\kappa_5^2\over (2\pi)^4}\int
d^4x^{\prime}d^4pe^{ip\cdot(x- x^{\prime})}
{[
J_2(qe^{b|w|}/b)  +i
Y_2(qe^{b|w|}/b)]
 \over q[J_1(q/b)+i
Y_1(q/b)]}S^{^{TT}}_{\mu\nu}(x^{\prime})
\nonumber \\
&&=-2\kappa_5^2\int d^4x^{\prime}G^{{\rm GKR}}(x,
x^{\prime},w,0) S^{^{TT}}_{\mu\nu}(x^{\prime})
\label{1.12}
\end{eqnarray}
(here $q^2=(p^0)^2-\bar{p}^2$) \cite{footnote3}. As constructed, the
propagator of Eq. (\ref{1.12}) is based on outgoing Hankel functions, and
is thus a natural candidate for causality. Whether or not $G^{{\rm
NM}}(x,0,w,0)$ is causal thus depends on determining whether or
not $G^{{\rm GKR}}(x,0,w,0)$ is indeed causal and then comparing the two.

When viewed as a function in the complex $p^0$ plane, $G^{{\rm GKR}}(x,
x^{\prime},w,0)$ is found possess both poles and cuts. Recalling that
$J_{1}(y)$, $J_{2}(y)$, $Y_{1}(y)$ and $Y_{2}(y)$ respectively behave as
$y/2$, $y^2/8$, $-2/\pi y+O(y)$ and $-4/\pi y^2-1/\pi$ near $y=0$, we see
that the integrand $[J_2(qe^{b|w|}/b)  +i Y_2(qe^{b|w|}/b)]/q[J_1(q/b)+i
Y_1(q/b)]$ behaves as $2be^{-2b|w|}/q^2$ near $q^2=0$, to thus precisely
generate a massless graviton pole term of exactly the same form as the
one in $G^{{\rm NM}}(x,0,w,0; {\rm SING})$ which is exhibited as the first
term in Eq. (\ref{1.9}). For the cut structure we recall that
$Y_2(qe^{b|w|}/b)$ and $Y_1(q/b)$ are both multiple-valued functions
with branch points at zero argument, viz. at $p^0=\pm
|\bar{p}|$. Calculation of the discontinuities across the associated
branch cuts is fairly lengthy, but is found to yield \cite{Mannheim2005}
none other than a KK continuum of terms of precisely the same form as the
ones in $G^{{\rm NM}}(x,0,w,0; {\rm SING})$ which are exhibited in Eq.
(\ref{1.9}) \cite{footnote4}. We thus conclude that $G^{{\rm GKR}}(x,
0,w,0; {\rm SING})$ and $G^{{\rm NM}}(x,0,w,0; {\rm SING})$ are identical
to each other, with the singular terms in $G^{{\rm GKR}}(x, 0,w,0; {\rm
SING})$ generating none other than the normalized mode contribution to
$G^{{\rm NM}}(x,0,w,0; {\rm SING})$. To complete the discussion we thus
need to compare the circle at infinity contributions to the
two propagators.

\section{Upper half plane determination of the GKR propagator}

Since we have taken the singularities of the $G^{{\rm GKR}}(x,0,w,0)$
propagator to all lie below the real $p^0$ axis, we can evaluate 
$G^{{\rm GKR}}(x,0,w,0)$ in two equivalent ways -- we can either
close the contour below the real $p^0$ axis and include both the
lower half plane circle at infinity and the singular $G^{{\rm GKR}}(x,
0,w,0; {\rm SING})$ term, or we can close the contour above where only
the upper half plane circle at infinity will then contribute. And as we
shall see, the equivalence of these two procedures will prove
instructive. In order to make the calculations simple enough to be
tractable but still rich enough to enable us to explore causal structure,
we shall take the source on the brane to be of a particularly simple and
convenient form, viz. we shall take it to be given by
$S^{^{TT}}_{\mu\nu}(x^{\prime})=A^{^{TT}}_{\mu\nu}\delta(t^{\prime})$ 
where
$A^{^{TT}}_{\mu\nu}$ is a constant TT tensor. With this choice for the
source Eq. (\ref{1.12}) simplifies to 
\begin{equation}
h^{^{TT}}_{\mu\nu}(x,|w|;{\rm GKR})=-{\kappa_5^2A^{^{TT}}_{\mu\nu}\over
2\pi}\int_{-\infty}^{+\infty} dp^0e^{-ip^0t}
{[
J_2(p^0e^{b|w|}/b)  +i
Y_2(p^0e^{b|w|}/b)]
 \over p^0[ J_1(p^0/b)+i
Y_1(p^0/b)]}~~.
\label{1.13}
\end{equation}
On the upper half circle we can set $p^0=Pe^{i\theta}$ where $P$ is
very large and $\theta$ lies in the range $0<\theta<\pi$. From the
standard behavior of the Bessel functions when their argument is large, we
find that the upper half circle circle at infinity
contribution to $h^{^{TT}}_{\mu\nu}(x,|w|;{\rm GKR})$ (as traversed
counter-clockwise) evaluates in leading order to
\begin{equation}
h^{^{TT}}_{\mu\nu}(x,|w|;{\rm UHPC})=-{\kappa_5^2A^{^{TT}}_{\mu\nu}\over
2\pi e^{b|w|/2}}\int_{0}^{\pi} id\theta e^{-iPe^{i\theta}\alpha}
(-i)\left[1-{15b \over 8iPe^{i\theta}e^{b|w|}}
+{3b \over 8iPe^{i\theta}}\right]
\label{1.14}
\end{equation}
where $\alpha$ is as given in Eq. (\ref{1.11}). With the integral in Eq.
(\ref{1.14}) being straightforward, on letting $P$ go to infinity, we
finally obtain
\begin{equation}
h^{^{TT}}_{\mu\nu}(x,|w|;{\rm GKR})=-h^{^{TT}}_{\mu\nu}(x,|w|;{\rm
UHPC})
={\kappa_5^2A^{^{TT}}_{\mu\nu}\over e^{b|w|/2}}\theta(\alpha)
\left[1+{15b\alpha \over 8e^{b|w|}}
-{3b\alpha \over 8} +O(\alpha^2)\right]
\label{1.15}
\end{equation}
to leading order. As we see, with the emergence of the overall
$\theta(\alpha)$ factor, $h^{^{TT}}_{\mu\nu}(x,|w|;{\rm GKR})$ does indeed
takes support only on and within the $AdS_5$ lightcone, with the
Giddings, Katz and Randall propagator thus being the causal one we
seek. 

\section{Lower half plane determination of the GKR propagator}

Analogously to the above, the contribution of the lower half $p^0$
plane circle at infinity (as traversed  clockwise) evaluates to
\begin{equation}
h^{^{TT}}_{\mu\nu}(x,|w|;{\rm
LHPC})=-{\kappa_5^2A^{^{TT}}_{\mu\nu}\over 2\pi
e^{b|w|/2}}\int_{\pi}^{0} id\theta e^{+iPe^{i\theta}\alpha}
(-1)i\left[1+{15b \over 8iPe^{i\theta}e^{b|w|}}
-{3b \over 8iPe^{i\theta}}\right]~~,
\label{1.16}
\end{equation}
with the full $h^{^{TT}}_{\mu\nu}(x,|w|;{\rm GKR})$ immediately evaluating
to
\begin{eqnarray}
h^{^{TT}}_{\mu\nu}(x,|w|;{\rm GKR})&=&h^{^{TT}}_{\mu\nu}(x,|w|;{\rm
SING})-h^{^{TT}}_{\mu\nu}(x,|w|;{\rm LHPC})
\nonumber \\
&=&h^{^{TT}}_{\mu\nu}(x,|w|;{\rm
SING})-{\kappa_5^2A^{^{TT}}_{\mu\nu}\over
e^{b|w|/2}}\theta(-\alpha)
\left[1+{15b\alpha \over 8e^{b|w|}}
-{3b\alpha \over 8} +O(\alpha^2)\right]~~.
\label{1.17}
\end{eqnarray}
Since $\theta(\alpha)+\theta(-\alpha)=1$, combining Eqs. (\ref{1.15})
and (\ref{1.17}) then shows that $h^{^{TT}}_{\mu\nu}(x,|w|;{\rm SING})$
is given by
\begin{equation}
h^{^{TT}}_{\mu\nu}(x,|w|;{\rm
SING})={\kappa_5^2A^{^{TT}}_{\mu\nu}\over e^{b|w|/2}}
\left[1+{15b\alpha \over 8e^{b|w|}}
-{3b\alpha \over 8} +O(\alpha^2)\right]~~,
\label{1.18}
\end{equation}
and that it is related to $h^{^{TT}}_{\mu\nu}(x,|w|;{\rm GKR})$ according
to 
\begin{equation}
h^{^{TT}}_{\mu\nu}(x,|w|;{\rm
GKR})=\theta(\alpha)h^{^{TT}}_{\mu\nu}(x,|w|;{\rm SING})~~.
\label{1.19}
\end{equation}
Consequently, we see that the singular 
$h^{^{TT}}_{\mu\nu}(x,|w|;{\rm SING})$ term takes support outside the
$AdS_5$ lightcone even though the full $h^{^{TT}}_{\mu\nu}(x,|w|;{\rm
GKR})$ itself does not, with it precisely being the contribution of the
circle at infinity which restores causality. Finally, comparing now with 
the normalized mode based propagator $G^{{\rm NM}}(x,0,w,0)$
discussed earlier, we see that the respective circle at infinity
contributions behave entirely differently; and since we had shown that 
$G^{{\rm GKR}}(x, 0,w,0; {\rm SING})$ and $G^{{\rm NM}}(x,0,w,0; {\rm
SING})$ are identical to each other, we can conclude that the normalized
mode based $G^{{\rm NM}}(x,0,w,0; {\rm SING})$ takes support outside the
$AdS_5$ lightcone, with the full $G^{{\rm NM}}(t>0,x=0, y=0, z=0,0,w,0)$
with $t>0$ then doing so too. It is thus the Hankel based propagator and
not the normalized mode based one which is the appropriate one
for the brane world.

\section{The difference between the NM and GKR propagators}

Even though both the ${\rm NM}$ and ${\rm GKR}$ propagators allow one to
integrate Eq. (\ref{1.2}), since the two propagators do not coincide with
each other, they must actually be solving Eq. (\ref{1.2}) in different
ways. In fact the difference lies in how they satisfy the junction
condition at the brane. Specifically, for the Giddings Katz and Randall
propagator, Eq. (\ref{1.2}) is found to decompose into two separate
equations, viz.
\begin{equation}
\left[{\partial^2 \over \partial |w|^2}
-4\left({d A\over
d|w|}\right)^2
+e^{-2A}\eta^{\alpha\beta}\partial_{\alpha}\partial_{\beta}
\right]h^{^{TT}}_{\mu\nu}(x,|w|;{\rm GKR}) =0~~,
\label{1.20}
\end{equation}
\begin{equation}
\delta(w)\left[{\partial \over
\partial|w|}-2{d A\over
d|w|}\right]h^{^{TT}}_{\mu\nu}(x,|w|;{\rm
GKR})=-\kappa_5^2\delta(w)S^{^{TT}}_{\mu\nu}~~.
\label{1.21}
\end{equation}
However, because the volcano potential modes obey the source-free
junction condition of Eq. (\ref{1.4}), their sum must do so too.
Consequently, the normalized mode based propagator must break up Eq.
(\ref{1.2}) in a different way, with explicit calculation revealing that
it does so according to
\begin{equation}
\frac{1}{2}\left[{\partial^2 \over \partial |w|^2}
-4\left({d A\over
d|w|}\right)^2
+e^{-2A}\eta^{\alpha\beta}\partial_{\alpha}\partial_{\beta}
\right]h^{^{TT}}_{\mu\nu}(x,|w|;{\rm NM})
=-\kappa_5^2\delta(w)S^{^{TT}}_{\mu\nu}~~,
\label{1.22}
\end{equation}
\begin{equation}
\delta(w)\left[{\partial \over
\partial|w|}-2{d A\over
d|w|}\right]h^{^{TT}}_{\mu\nu}(x,|w|;{\rm NM})=0
\label{1.23}
\end{equation}
instead. Comparing Eqs. (\ref{1.20}) and (\ref{1.21})  with
Eqs. (\ref{1.22}) and (\ref{1.23}), we thus see that working
with modes which obey the source-free Eqs. (\ref{1.3}) and (\ref{1.4})
simply fails to capture the proper junction condition structure of the
theory when the source is present \cite{footnote5}.
Acknowledgment:
This work grew out of a study of brane-world fluctuations in which 
the author was engaged with Dr. A. H. Guth, Dr. D. I. Kaiser and Dr. A.
Nayeri, and the author would like to thank them for their many
helpful comments.


\begin{thebibliography}{}
\bibitem{footnote2} The $m^2=0$ graviton is bound state normalized and the
KK continuum modes are normalized  as box normalized
Bessel functions, viz. as $2\int_1^Ldx
xf_{m_i}(m_ix/b)f_{m_j}(m_jx/b)=\delta_{ij}L/\pi$ where $x=e^{b|w|}$.
\bibitem{footnote3} That Eq. (\ref{1.12}) is a solution to
Eq. (\ref{1.2}) is checked via the use of Eq. (\ref{1.3})
(which both $J_2(qe^{b|w|}/b)$ and 
$Y_2(qe^{b|w|}/b)$ identically satisfy with $q^2$ replacing $m^2$) and the
Bessel function relation given in Ref.
\cite{footnote1}.
\bibitem{footnote4} Contact with Eq. (\ref{1.9}) is made since the
imaginary part of $[J_2(qe^{b|w|}/b)  +i Y_2(qe^{b|w|}/b)]/[J_1(q/b)+i
Y_1(q/b)]$ is precisely $2[J_1(q/b)Y_2(qe^{b|w|}/b)-
Y_1(q/b)J_2(qe^{b|w|}/b)]/[J_1^2(q/b)+Y_1^2(q/b)]$.
\end{thebibliography}
\end{document}